\documentclass[12pt]{article}

\usepackage{amsmath}
\usepackage{esvect} 
\usepackage{tikz}
\usetikzlibrary{arrows}
\usetikzlibrary{positioning}
\usepackage{amsfonts}
\usepackage{cite}
\usepackage[margin=1in]{geometry}
\usepackage{amsthm}

\usepackage{epic}
\usepackage{mathtools}
\usepackage{url}
\usepackage{hyperref}
\usetikzlibrary{calc,patterns,angles,quotes}
\usepackage{graphicx}
\usepackage{float}
\usepackage{amssymb}
\usepackage{caption}
\usepackage{subcaption}
\usepackage{xcolor}
\usepackage{bm}
\usepackage{multicol} 
\usepackage{seqsplit}
\usepackage{mathrsfs}
\usepackage{comment}
\usetikzlibrary{decorations.pathmorphing}
\usepackage[affil-it]{authblk}
\usepackage{tikz-cd}
\numberwithin{equation}{section}
\usepackage{braket}
\allowdisplaybreaks[4]

\def\r{\rho}                                     
\def\t{\tau}

    \let\r=\rho
 \let\t=\tau

\def\nn{\nonumber} \def\bd{\begin{document}} \def\ed{\end{document}}
\def\ds{\documentstyle} \let\fr=\frac \let\bl=\bigl \let\br=\bigr
\let\Br=\Bigr \let\Bl=\Bigl
\let\bm=\bibitem
\let\na=\nabla

\newcommand{\be}{\begin{equation}}
\newcommand{\ee}{\end{equation}}

\newcommand{\bea}{\begin{eqnarray}}
\newcommand{\eea}{\end{eqnarray}}

\usetikzlibrary{decorations.pathmorphing}
\usetikzlibrary{arrows.meta}
\usetikzlibrary{decorations.markings}

\definecolor{darkpastelgreen}{rgb}{0.01, 0.75, 0.24 }
\definecolor{hooker\'sgreen}{rgb}{0.0, 0.44, 0.0}
\definecolor{indiagreen}{rgb}{0.07, 0.53, 0.03}
\definecolor{islamicgreen}{rgb}{0.0, 0.56, 0.0}

\begin{document}
\title{\textbf{TT deformations in general dimensions} }

\author{\textbf{\small{ Marika Taylor}}}
\date{}
\affil{\small{Mathematical Sciences and STAG Research Centre, University of Southampton, Highfield, Southampton SO17 1BJ, United Kingdom}}
\maketitle

\begin{center}
\textbf{Abstract} 
\end{center}

It has recently been proposed that Zamoldchikov's $T \bar{T}$ deformation of two-dimensional CFTs describes the holographic theory dual to AdS$_3$ at finite radius. In this note we use the Gauss-Codazzi form of the Einstein equations to derive a relationship in general dimensions between the trace of the quasi-local stress tensor and a specific quadratic combination of this stress tensor, on constant radius slices of AdS. We use this relation to propose a generalization of Zamoldchikov's $T \bar{T}$ deformation to conformal field theories in general dimensions. This operator is quadratic in the stress tensor and retains many but not all of the features of  $T \bar{T}$. To describe gravity with gauge or scalar fields, the deforming operator needs to be modified to include appropriate terms involving the corresponding R currents and scalar operators and we can again use the Gauss-Codazzi form of the Einstein equations to deduce the forms of the deforming operators. We conclude by discussing the relation of the quadratic stress tensor deformation to the stress energy tensor trace constraint in holographic theories dual to vacuum Einstein gravity.  
 

\newpage


\section{Introduction}

A decade ago, Zamolochikov \cite{Zamolodchikov:2004ce} explored deformations of two-dimensional conformal field theories by an operator that is quadratic in the stress energy tensor, called the $T \bar{T}$ operator. This operator is defined as a bilocal operator, 
\be
T \bar{T} (x,y) = T^{ij} (x) T_{ij} (y) - T^{i}_{i} (x) T^{j}_{j} (y)
\ee
where $T_{ij}$ is the stress energy tensor. In a two-dimensional CFT this operator was shown by Zamoldchikov to have a remarkable OPE structure as $x \rightarrow y$:
\be
T \bar{T} (x,y) = {\cal T} (y) + \sum_{\alpha} A_{\alpha} (x-y) \nabla_y {\cal O}_{\alpha} (x). \label{one1}
\ee
Here ${\cal O}_{\alpha}$ denote local operators and the function $A_{\alpha} (x-y)$ can be divergent as $x \rightarrow y$; this relation implies that we can identify $T \bar{T}$ as a local operator ${\cal T}(y)$, modulo derivatives of other local operators. 
The $T \bar{T}$ operator can be used to deform the conformal field theory, generating a family of theories characterized by the coupling of this operator. While the deforming operator is irrelevant, its particular properties imply that the resulting theory is more predictive than a generic non-renormalizable quantum field theory. 

It has recently been proposed that the $T \bar{T}$ deformation is relevant to understanding the holographic theory dual to AdS$_3$ with a finite radial cutoff \cite{McGough:2016lol}. The basic idea is that the holographic theory at finite radius is a member of the family of deformed theories discussed by Zamolodchikov. These ideas were explored further in a number of other works, see \cite{Jensen:2017tnb,Guica:2017lia,Giribet:2017imm,Kraus:2018xrn,Cottrell:2018skz,Aharony:2018vux,Bzowski:2018pcy,Dubovsky:2018bmo,Chakraborty:2018kpr}. 

There have been a number of previous attempts to set up a holographic correspondence at finite radius, using various approaches. In the early days of AdS/CFT, the role of the holographic renormalization group was explored, with \cite{deBoer:1999tgo} relating radial flow to renormalization group. In what follows we will use the sharp dictionary between radial Hamiltonian evolution and holographic operators developed in \cite{Papadimitriou2004}. In \cite{Heemskerk:2009pn,Heemskerk:2010hk}, it was proposed that the holographic dual at finite radius should be interpreted as a deformed CFT.

Another approach to understanding holography at finite radius follows from the fluid gravity correspondence. In \cite{Brattan:2011my}, the fluids dual to finite cutoff surfaces in asymptotically AdS black brane geometries were analysed, and it was argued that these fluids should be related to fluids in appropriately deformed conformal field theories. Note that one can set up a Dirichlet problem at finite radius not just in asymptotically AdS geometries, but also in Ricci flat spacetimes \cite{Bredberg:2010ky,Bredberg:2011jq,Compere:2011dx,Compere:2012mt}. In \cite{Compere:2011dx,Compere:2012mt}, the corresponding Ward identities for the fluid stress tensor were used to infer information about the putative holographic quantum field theory dual to such spacetimes.

\bigskip

The properties of the $T \bar{T}$ operator are very specific to two dimensions. Potential analogues of the $T \bar{T}$ operator in higher dimensions have been discussed from the field theory perspective in \cite{Cardy:2018sdv,Bonelli:2018kik}. In particularly, Cardy proposed in  \cite{Cardy:2018sdv} that a deformation involving the square root of the determinant of the stress tensor may be an appropriate generalisation. This possibility was further explored in \cite{Bonelli:2018kik}, although \cite{Bonelli:2018kik} also mentions that a quadratic generalization to dimensions higher than two may be more natural holographically. 

In this short note we use holography to propose a generalisation of the $T \bar{T}$ deformation to dimensions higher than two. Following the approach of \cite{Kraus:2018xrn}, we derive an expression for the Ward identity involving the trace of the stress tensor, at finite radius. This analysis implies that in a $d$-dimensional theory we should consider the operator
\be
{\cal T} = T^{ij} T_{ij}  - \frac{1}{(d-1)} T^{i}_{i}  T^{j}_{j}. \label{res}
\ee
This clearly reduces to the $T \bar{T}$ operator in two dimensions. In general dimensions the operator does not share all the special properties uncovered by Zamolodchikov in two dimensions, but we argue that these properties are not required if one is only considering states that are stationary and spatially homogeneous. We demonstrate that the energy relation for constant radius hypersurfaces in black branes indeed precisely matches the energy relation obtained by deforming a conformal field theory with this operator. 

As highlighted in \cite{Guica:2017lia,Kraus:2018xrn,Bzowski:2018pcy}, the $T \bar{T}$ operator is not sufficient to describe AdS$_3$ gravity with matter: for example, additional deformations are required if one includes bulk scalars and gauge fields. By deriving the Ward identity for the trace of the stress of the stress tensor in the presence of gauge fields, we propose the required deformation involving R currents in the dual field theory. We also consider briefly the case of scalar fields. 

The quadratic combination of the stress tensor in \eqref{res} has appeared in earlier literature: the vacuum Einstein equations (with zero cosmological constant) force the induced Brown York tensor on a hypersurface to satisfy the constraint ${\cal T} = 0$ \cite{Compere:2011dx,Compere:2012mt}. In section \ref{sec:seven} we explain the relationship between the current work and this constraint; it is the Gauss-Codazzi relations (in particular, the Hamiltonian constraint) on a constant radius surface that picks out the combination \eqref{res} in both cases. 

\bigskip

The plan of this paper is as follows. In section \ref{sec:two} we consider AdS gravity in general dimensions and show that the trace identity for the stress energy tensor at finite radius implies a deformation by an operator of the form \eqref{res}. In section \ref{sec:three} we consider AdS branes and derive an expression for the effective energy as a function of the cutoff radius. We consider the dual interpretation of the operator in section \ref{sec:four}. In sections \ref{sec:five} and \ref{sec:six} we consider generalisations to include gauge and scalar fields in the bulk. We discuss the relation of the TT deformation to the defining holographic relation for holographic theories dual to vacuum Einstein gravity in section \ref{sec:sixa}. We conclude in section \ref{sec:seven}. 

\section{AdS gravity} \label{sec:two}

The bulk (Euclidean) action for pure AdS gravity is
\be
I_{M} = - \frac{1}{16 \pi G} \int d^{d+1} x \sqrt{g} \left ( R + d (d-1) \right ),
\ee
where we set the AdS radius to one. The boundary terms for this action include the standard Gibbons-Hawking-York term and the counterterms derived in \cite{DeHaro2001}:
\be
I_{\partial M} = - \frac{1}{8 \pi G} \int d^d x \sqrt{h} \left ( K -  (d-1) + \cdots \right ) 
\ee
where $K$ is the trace of the extrinsic curvature and the ellipses denote terms involving the intrinsic curvature of the metric $h$. 

We now introduce a coordinate system such that the metric takes the form
\be
ds^2  = dr^2 + \gamma_{ij} (x,r) dx^i dx^j \label{fol}
\ee
in terms of which the extrinsic curvature takes the simple form
\be
K_{ij} = \frac{1}{2} \partial_r \gamma_{ij}. 
\ee
As exploited in the Hamiltonian renormalization approach of \cite{Papadimitriou2004}, one can use the standard Gauss-Codazzi relations to rewrite the action as 
\be
I = - \frac{1}{16 \pi G} \int d^{d+1} x \sqrt{g} \left ( {\cal R} + K^2 - K^{ij} K_{ij} + d (d-1) \right ) + \frac{1}{8 \pi G} \int d^{d}x \sqrt{h} \left ( (d-1) + \cdots \right )
\ee
where ${\cal R}$ denotes the Ricci scalar of constant radius hypersurfaces and the ellipses again denote terms involving the intrinsic curvature of the boundary metric. 

One can use the Gauss-Codazzi equations to express the Einstein equations in the coordinate system \eqref{fol}. For us, the most relevant equation is the $(r r )$ equation which implies
\be
K^2 - K^{ij} K_{ij} = {\cal R} + d (d-1). \label{gc}
\ee
We now define the quasi-local stress energy tensor as 
\be
\delta I = \frac{1}{2} \int d^{d} x \sqrt{h} T^{ij}  \delta h_{ij}
\ee
which gives
\be
T_{ij} = \frac{1}{8 \pi G} \left ( K_{ij} - K \gamma_{ij} +  (d-1) \gamma_{ij} + \cdots \right )
\ee
The first two terms are the Brown-York quasi-local stress tensor. The third term follows from the leading order counterterm i.e. it contains information about the renormalization required to compute quantities in the conformal field theory. While one could add any term proportional to the metric to the Brown-York tensor without spoiling the conservation of the stress tensor, the specific coefficient arising here follows from holographic renormalization \cite{Balasubramanian1999a,DeHaro2001}. 

Now let us restrict to slices for which the intrinsic curvature vanishes i.e. the terms in ellipses are zero. Then
\be
T_{i}^{i} = \frac{(d-1)}{8 \pi G} (d- K)
\ee
We can also show that 
\be
T^{ij} T_{ij} = \frac{(d-1)}{64 \pi^2 G^2} \left ( K^2 - 2 (d-1) K + d  (d-2) \right )
\ee
where we use the Gauss-Codazzi relation to eliminate $K^{ij} K_{ij}$. Then
\be
- 4 \pi G \left ( T_{ij} T^{ij} - \frac{1}{(d-1)} (T^i_i)^2 \right ) = \frac{(d-1)}{8 \pi G} (d-K) = T^{i}_i.
\ee
Again we highlight the role of the counterterm (and the corresponding term linear in the metric in the stress tensor) in determining this expression. 

This calculation suggests that in general dimensions the deformation that one should consider in general dimensions is
\be
{\cal T} \equiv \left ( T_{ij} T^{ij} - \frac{1}{(d-1)} (T^i_i)^2 \right )
\ee
which manifestly reduces to the $T \bar{T}$ deformation in $d=2$. We can then write the trace relation as 
\be
T^{i}_{i} = - \lambda {\cal T}  \label{trace}
\ee
where $\lambda = 4 \pi G$. 

\section{Energy spectrum} \label{sec:three}

In this section we consider asymptotically AdS black branes in general dimensions and show how their energy (as defined by the quasi-local stress tensor) changes with the radius. In the following section we will show how such a relation follows from interpreting the cutoff geometries in terms of theories deformed by the operator ${\cal T}$. 

\bigskip

Let us consider a general static metric of the form
\be
ds^2 = \rho^2 f(\rho)^2 d\t^2 + \frac{d\rho^2}{\rho^2 f(\rho)^2} + \rho^2 dx^a dx_a. \label{static1}
\ee
A metric of the form \eqref{static1} has the following Ricci tensor:
\begin{eqnarray}
R_{ \hat{0} \hat{0}} &=& \left ( - d f^2 - (d+1) \r f \partial_{\r} f - \partial_{\r} (\r^2 f \partial_{\r} f ) \right ) \label{curv} \\
R_{ \hat{1} \hat{1}} &=& \left ( - d f^2 - (d+1)  \r f \partial_{\r} f - \partial_{\r} (\r^2 f \partial_{\r} f ) \right ) \nonumber \\
R_{ \hat{a} \hat{b}} &=& \left ( - d f^2 - 2\r f \partial_{\r} f \right ) \delta_{\hat{a} \hat{b}} \nonumber 
\end{eqnarray}
where we have introduced an orthonormal frame
\be
e^{\hat{0}} = \r f(\r) d\t; \qquad 
e^{\hat{1}} = \frac{d\r}{\r f(\r)}; \qquad
e^{\hat{a}} = \r d x^{a}.
\ee
The extrinsic curvature is 
\begin{eqnarray}
K_{\t \t} &=& \frac{1}{2}  f(\r) \partial_{\r} \left ( f(\r)^2  \right ) \\
K_{ab} &=& f(\r) \r \delta_{ab}. \nn
\end{eqnarray}
The quasi-local stress energy tensor implies that
\be
T_{\t \t} = \frac{1}{8 \pi G} \left ( D  - \frac{D}{\r} f(\r) \right ),
\ee
where $D$ is the number of spatial dimensions. The quasi-local energy is then 
\be
{\cal E} = \frac{1}{8 \pi G} \int d^{D} x \r^{D} \left ( D - \frac{D}{\r} f(\r) \right ) = \frac{D V_D \r^D}{8 \pi G}  \left (1 - \frac{1}{\r} f(\r) \right ), \label{quasi}
\ee
where $V_D$ is the regulated (dimensionless) volume of the spatial directions. 

For AdS-Schwarzschild (with flat horizon) in $(d+1)$ dimensions
\be
f(\r)^2 = \r^2 - \frac{\mu}{\r^{D-1}}
\ee 
and thus
\be
{\cal E} = \frac{D V_D \r^D}{8 \pi G}  \left (1 - \left ( 1 - \frac{\mu}{\r^{D+1}} \right )^{\frac{1}{2}} \right ).
\ee
We can then express the dimensional ratio 
\be
\epsilon =  {\cal E} \r = \frac{D V_D \r^{D+1}}{8 \pi G}  \left (1 - \left ( 1 - \frac{\mu}{\r^{D+1}} \right )^{\frac{1}{2}} \right )
\ee
as
\be
\epsilon  = \frac{D V_D \r^{d}}{2 \lambda}  \left (1 - \left ( 1 - \frac{\lambda M}{\r^{d}} \right )^{\frac{1}{2}} \right ) \label{par}
\ee
in terms of the (dimensionless) mass parameter $4 \pi M = \mu/G$ and $d = D+1$. In the limit of $\r \rightarrow \infty$ this expression gives the finite renormalized quantity
\be
\epsilon \rightarrow \frac{D V_D}{4} M.
\ee
The relation \eqref{quasi} clearly holds for any static metric with flat spatial sections. In particular, we can apply the relation to well-known geometries such as charged AdS black branes for which
\be
f(\r)^2 =  \r^2 - \frac{\mu}{\r^{D-1}} + \frac{Q^2}{\r^{2 (D-1)}}
\ee
with Q the electric charge, and thus 
\be
\epsilon =  {\cal E} \r = \frac{D V_D \r^{D+1}}{2 \lambda}  \left (1 - \left ( 1 - \frac{\lambda M}{\r^{D+1}} + \frac{Q^2}{\r^{2D}} \right )^{\frac{1}{2}} \right ) \label{charg-rel}
\ee
In the limit as $\r \rightarrow \infty$ this again tends to 
\be
\epsilon = \frac{1}{4} D V_D M,
\ee
i.e. the charge parameter drops out of the energy. We will return to charged black branes below.

\section{CFT deformation} \label{sec:four}

Now let us turn to the interpretation in terms of CFT deformations. In this section we consider the behaviour of the operator
\be
{\cal T} \equiv \left ( T_{ij} T^{ij} - \frac{1}{(d-1)} (T^i_i)^2 \right ) \label{curlt}
\ee
in general dimensions. The key point is to analyse whether there is indeed an appropriately defined composite operator i.e. that one can define
\be
{\cal T} (x) = {\rm Lim}_{x \rightarrow y} \left ( T^{ij} (x) T_{ij} (y) - \frac{1}{(d-1)} T^{i}_i (x) T^{j}_j (y) \right ) \label{limit-com}
\ee
 There is a fundamental difference between the behaviour of this operator in two dimensions and the behaviour for $d > 2$. 
 
To illustrate this, we first consider the expectation value of the right hand side in the conformal vacuum. 
In a $d$-dimensional CFT, the two point function of the stress tensor at separated points is given by
\be
\langle T_{ij} (x) T_{kl} (0) \rangle = \frac{1}{ x^{2d}} \left ( \frac{1}{2} \left ( I_{i k} I_{j l} + I_{il} I_{jk} \right ) - \frac{1}{d} \delta_{ij} \delta_{kl} \right )
\ee
where 
\be
I_{ij} = \delta_{ij} - \frac{2 x_i x_j}{x^2}
\ee
and implicitly we work in Euclidean signature. It is then straightforward to show that
\be
\langle T_{ij} (x) T^{ij} (0) \rangle = \frac{(d-2)}{x^{2d}} \label{sing}
\ee
and
\be
\langle T_{i}^i (x) T^{j}_j (0) \rangle = 0.
\ee
Thus for a CFT in the conformal vacuum the combination
\be
 \langle T^{ij} (x) T_{ij} (y) \rangle - \frac{1}{(d-1)} \langle T^{i}_i (x) T^{j}_j (y) \rangle = \frac{(d-2)}{| x - y |^{2d}} \label{sing2}
\ee
is only position independent in $d=2$. 

Note that, using standard regularisation techniques, one can renormalize so that \eqref{sing} is well defined in the distributional sense as $x \rightarrow 0$. For example, using differential regularisation, for $d$ odd one can write
\be
\frac{1}{x^{2d}} \sim \Box^{\frac{d}{2} + 1} \left ( \frac{1}{x^{d-2}} \right ),
\ee
while for $d$ even 
\be
\frac{1}{x^{2d}} \sim \Box^{\frac{d}{2} + 1} \left ( \frac{\log (x^2 \mu^2)}{x^{d-2}} \right ),
\ee
where $\mu$ is an arbitrary (renormalization scale) parameter. 

\bigskip

The arguments of Zamolodchikov \cite{Zamolodchikov:2004ce} in $d=2$ do not rely on conformal symmetry - but on local translational and rotational symmetry, which in particular imply that the stress energy tensor is symmetric and conserved. In two dimensions, these properties suffice to show that the operator product expansion for the quadratic $T \bar{T}$ combination of the stress tensor is of the form
\be
T^{ij} (z) T_{ij} (z') - T^{i}_i (z) T^{j}_j (z') = \sum_{\alpha} A_{\alpha} (z - z') {\cal O}_{\alpha} (z') \label{ope1}
\ee
where the coefficients $A_{\alpha}$ are coordinate independent unless the operator ${\cal O}_{\alpha}$ is itself the coordinate derivative of another local operator. 
This form for the OPE thus implies that we can identify the $T \bar{T}$ operator as a local operator ${\cal T}$, modulo derivatives of local operators, as stated in \eqref{one1}. 

It is instructive to understand how Zamolodchikov's arguments change in dimensions different from two. The OPE still takes the form \eqref{ope1} in all dimensions. However, in dimensions greater than two, the conservation of the stress tensor places fewer restrictions on the right hand side of \eqref{ope1}; the coefficients can be divergent even when ${\cal O}_{\alpha}$ is not the derivative of a local operator.  This is already apparent from the expression \eqref{sing2}. 

These issues do not mean that one cannot define a composite local operator ${\cal T}$ in dimensions higher than two; the definition however depends on the renormalization procedure (and hence on the full short distance behaviour of the theory). This is significant, as we cannot assume that the UV theory is conformal: we instead want to consider a family of theories characterized by irrelevant deformations of a CFT.  
As we discuss below, for matching the energy spectrum with the holographic description, we can however avoid discussing details of the regularisation procedure, as the states under consideration are static and translationally invariant. 

\subsection{Energy spectrum}

In \eqref{par}, we found an energy spectrum within gravity which can be expressed in the form
\be
{\epsilon} = \frac{C}{\alpha} \left ( 1 - \left ( 1 - n \alpha \right )^{\frac{1}{2}} \right ),
\ee
where $C$ is a constant, $\alpha$ is a dimensionless coupling and $n$ is a dimensionless parameter characterising the energy (eigen)state. 
The constant $C$ is the overall normalization constant for the energy in the undeformed theory i.e. 
\be
\epsilon(0) = \frac{1}{2} C n. 
\ee
Following the analysis of \cite{Kraus:2018xrn}, one can observe that this quantity satisfies a non-linear differential equation
\be
\partial_{\alpha } {\cal \epsilon} = \frac{1}{2 C} {\epsilon} \left ( {\epsilon} + 2 \alpha \frac{d {\epsilon}}{d \alpha} \right ), \label{ode}
\ee
for any value of $n$. 

\bigskip

In two dimensions, this equation can be interpreted directly in terms of the defining Ward identity for the family of deformed theories. Let us consider the Euclidean theory on a cylinder of circumference $R$. A stationary state $| n \rangle $  has energy $E_n (R, \lambda)$, where the parameter $\lambda$ denotes the coupling of the operator ${\cal T}$, and momentum along the circle direction $P_n = 2 \pi l_n /R$, where $l_n$ is an integer. To match with the gravity analysis above, we will restrict to zero momentum states. 

Then, using the fact that ${\cal T}$ is well-defined as a local operator in two dimensions, Zamolodchikov \cite{Zamolodchikov:2004ce} showed that in the state $| n \rangle$ the expectation value of ${\cal T}$ is
\be
\langle n | {\cal T} | n \rangle = - \frac{2 }{R} E_n \frac{\partial E_n}{\partial R}. \label{tvev}
\ee
This follows from
\be
{\cal T} = 2 \left ( T_{\tau x} T_{\tau x} - T_{\tau \tau} T_{xx} \right ),
\ee
where we use coordinates $(\tau,x)$, and from the definition of the energy momentum tensor i.e.
\be
\langle n | T_{\tau \tau} | n \rangle = - \frac{E_n}{R}  \qquad
\langle n | T_{x x}  | n \rangle = - \frac{\partial E_n}{\partial R}
\ee
with 
\be
\langle n | T_{\tau x} | n \rangle = \frac{i}{R} P_n
\ee
when the momentum is non-zero. (Note that the minus sign in the first expression and the factor of $i$ in the final expression follow from the Euclidean signature.)

When we calculate an expression such as 
\be
\langle n  | T_{\tau \tau} T_{xx}  |  n \rangle 
\ee
we can always use a spectral decomposition
\be
\sum_{m} \langle n  | T_{\tau \tau} | m \rangle \langle m | T_{xx}   |  n \rangle e^{(E_n - E_m) \tau} e^{i (P_n - P_m) x} 
\ee
The proof that the operator ${\cal T}$ is independent of position (in $d=2$) implies that terms with $m \neq n$ must cancel and thus the expectation value factorises
\be
\langle n | {\cal T} | n \rangle = 2 \langle n | T_{\tau x} | n \rangle^2 - 2  \langle n | T_{\tau \tau} | n \rangle \langle n | T_{x x}  | n \rangle
\ee
leading to the result above \eqref{tvev} when the momentum vanishes. 

\bigskip

The defining relation for the family of QFTs is 
\be
\frac{d S}{d \lambda} = \int d^2 z \; {\cal T} (z),
\ee
where we define $S(0)$ as the CFT, which in turn implies that \cite{Zamolodchikov:2004ce,Smirnov:2016lqw}
\be
\frac{\partial E_n}{\partial \lambda} + 2 E_n \frac{\partial E_n}{\partial R} = 0. 
\ee
This is not yet comparable to \eqref{ode}, as the latter is expressed in terms of a dimensionless coupling. Letting 
\be
\alpha = \frac{\lambda}{R^2}
\ee
and using dimensional analysis to write the energy as 
\be
E_n = \frac{1}{R} \epsilon_n (\alpha),
\ee
where $\epsilon_n$ is dimensionless, we obtain 
\be
\partial_{\alpha } {\epsilon}_n = 2 {\epsilon}_n \left ( {\epsilon}_n + 2 \alpha \partial_{\alpha} \epsilon_n \right ), \label{ode2}
\ee
which indeed agrees with \eqref{ode}. 

\bigskip

Now let us consider the generalization to a theory in $d$ dimensions compactified on a symmetric torus of volume $V = R^D$. Repeating the two-dimensional arguments, let a stationary state $| n \rangle $  have energy $E_n (R, \lambda)$, where the parameter $\lambda$ denotes the coupling of the operator ${\cal T}$, and the momenta along the circle directions are $P^a_n = 2 \pi l^a_n /R$, where $l^a_n$ are integers. To match with the gravity analysis above, we will restrict to zero momentum states, although the generalization to include momentum would be straightforward. 

We can then calculate the expectation value of the operator ${\cal T}$ as  
\begin{eqnarray}
\langle n | \left ( T^{ij} T_{ij} - \frac{1}{D} (T^i_i)^2 \right ) | n \rangle &=& \sum_m \langle n | T^{ij} | m \rangle \langle m | T_{ij} | n \rangle e^{ (E_n - E_m) \tau} e^{i (P^a_n - P^a_m) x^a}   \\
&& \qquad - \frac{1}{D} \sum_m \langle n | T^{i}_i | m \rangle \langle m | T^{i}_i | n \rangle e^{ (E_n - E_m) \tau} e^{i (P^a_n - P^a_m) x^a}, \nonumber
\end{eqnarray}
where $\{ | m \rangle \}$ denotes a non-degenerate complete set of energy and momentum states. Since $|n \rangle$ itself is such an eigenstate, the superposition above immediately collapses onto the $n = m$ terms i.e. 
\be
\langle n | \left ( T^{ij} T_{ij} - \frac{1}{D} (T^i_i)^2 \right ) | n \rangle =  \langle n | T^{ij} | n \rangle \langle n | T_{ij} | n \rangle  - \frac{1}{D}  \langle n | T^{i}_i | n \rangle^2.  \label{def3}
\ee
Let us now consider the assumptions that go into this expression. The operator product expansion \eqref{ope1} in $d \neq 2$ implies that the two point function evaluated in a generic state is manifestly dependent on the points $z$ and $z'$; thus the expectation value of the operator ${\cal T}$ would in general depend on how one regularises the operators as they approach each other. 

On the other hand, since the original theory is a CFT and the deformation does not change its long distance behaviour, factorization at infinite separations remains applicable for any value of the coupling $\lambda$:
\be
 {\cal L}_{y \rightarrow \infty} \langle {\cal O}_{\alpha} (z + y) {\cal O}_{\beta} (z') \rangle = \langle {\cal O}_{\alpha} \rangle \langle {\cal O}_{\beta} \rangle
\ee 
where here $y$ can denote separation in time or space. 

The computation above assumes that the regularisation procedure does not affect the expectation value of ${\cal T}$ in an energy/momentum eigenstate i.e. it can effectively be computed by factorization. In particular, denoting the conformal vacuum of the CFT as $| 0 \rangle$, then 
we showed in \eqref{sing2} that the two point function of $TT$ needs to be regularised as the operators approach each other. The regularisation is such that 
\be
\langle 0 | {\cal T} | 0 \rangle =  \langle 0 | T^{ij} | 0 \rangle \langle 0  | T_{ij} | 0 \rangle  - \frac{1}{D}  \langle 0 | T^{i}_i | 0 \rangle^2 = 0,
\ee
where $\langle 0 | T_{ij} | 0 \rangle$ for the conformal vacuum (on the Euclidean plane). 

\bigskip

Now let us evaluate \eqref{def3} for finite energy states. 
The assumption that the state $| n \rangle$ is both static and spatially isotropic implies that
\be
\langle n | T_{\tau a} | n \rangle = 0 \qquad \langle n | T_{ab} | n \rangle = 0 \qquad a \neq b 
\ee
and
\be
\langle n | T_{\tau \tau} | n \rangle = - \frac{E_n}{V}
\ee
while for the stress tensor in the spatial directions
\be
\langle n | T_{ab} | n \rangle = \delta_{ab} \frac{P_n}{V}
\ee
where $P_n$ is the pressure. It is then straightforward to show that 
\be
\langle n | {\cal T} | n \rangle = \left ( 1 - \frac{1}{D} \right ) \frac{E_n^2}{V^2} + \frac{2 E_n P_n}{V^2}.
\ee
Note that the terms involving $P_n^2$ cancel between the $T^{ij} T_{ij}$ and $(T^i_i)^2$ combinations i.e. the relative factor between these terms guarantees this cancellation. It is due to this cancellation that the differential equation we obtain below is first order rather than second order. The pressure is related to the gradient of the energy as 
\be
P_n = - V \frac{\partial E_n}{\partial V}. 
\ee
Now, following the same logic as in two dimensions, the defining relation for the family of quantum field theories characterised by the coupling $\lambda$ is
\be
\frac{\partial E_n}{\partial \lambda} = \left ( 1 - \frac{1}{D} \right ) \frac{E_n^2}{V} - 2 E_n \frac{\partial E_n}{\partial V}
\ee
The dimensionless coupling $\alpha = \lambda/V^{1 + \frac{1}{D}}$ and by standard dimensional analysis 
\be
E_n = \frac{1}{V^{\frac{1}{D}}} \epsilon_n (\alpha)
\ee
where $\epsilon$ is dimensionless. Thus in $D$ spatial dimensions we find
\be
\partial_{\alpha} \epsilon_n =  \left ( 1 + \frac{1}{D} \right ) \left ( \epsilon_n^2 + 2 \alpha \epsilon_n \partial_{\alpha} \epsilon_n \right ),
\ee
which agrees with \eqref{ode} i.e. the energy spectrum found in gravity is indeed consistent with that of the family of deformed theories. 

\section{Charged black branes} \label{sec:five}

The generalization to include charge on the black holes is straightforward. We couple gravity to a gauge field via
\be
I_M = - \frac{1}{16 \pi G} \int d^{d+1} x \sqrt{g} \left (R - \frac{1}{4} F^{\mu \nu} F_{\mu \nu} + d (d-1) \right )
\ee
We can always introduce a coordinate system (locally) in which the metric takes the same form as before i.e.
\be
ds^2 = dr^2 + \gamma_{ij}(x,r) dx^i dx^j.
\ee
Using the Gauss-Codazzi relations we can then write the Einstein equations as 
\begin{eqnarray}
K^2 - K^{ij} K_{ij} &=& {\cal R} + d (d-1)  + \frac{1}{2} \dot{A}_i \dot{A}^i - \frac{1}{4} {\cal F}_{ij} {\cal F}^{ij} \\
\nabla_{i} K^{i}_j - \nabla_i K &=& \frac{1}{8} \dot{A}^i {\cal F}_{ij} \nonumber \\
\dot{K}^{i}_j  + K K^{i}_j &=& {\cal R}^{i}_j - \frac{1}{2} \dot{A}^i \dot{A}_j - \frac{1}{2 (1-d)} (\dot{A}^k \dot{A}_k) \delta^i_j \nonumber \\
&&  - \frac{1}{2} {\cal F}^{ik} {\cal F}_{jk} - \frac{1}{4 (1-d)} {\cal F}_{kl} {\cal F}^{kl} \delta^{i}_j - d \delta^i_j \nonumber
\end{eqnarray}
where the dots indicate radial derivatives. Here we have chosen a radial gauge for the gauge field, $A_r = 0$, and  
${\cal F}_{ij} = \partial_i A_j - \partial_j A_i$. The gauge field equations are then 
\be
\ddot{A}^i + K \dot{A}^i + \nabla_j {\cal F}^{ji}  = 0 \qquad
\nabla_{i} \dot{A}^i = 0 
\ee
Now let us restrict to the case where the curvature of constant radius slices is zero and the gauge field depends only on the radius. 

The quasi-local stress energy tensor is again
\be
T_{ij} = \frac{1}{8 \pi G} \left ( K_{ij} - K \gamma_{ij} +  (d-1) \gamma_{ij} \right )
\ee
and thus
\be
T_{i}^{i} = \frac{(d-1)}{8 \pi G} (d- K)
\ee
We can also show that 
\be
T^{ij} T_{ij} = \frac{1}{64 \pi^2 G^2} \left ( (d-1) ( K^2 - 2 (d-1) K + d  (d-2) ) - \frac{1}{2} \dot{A}_i \dot{A}^i  \right )
\ee
where we use the Gauss-Codazzi relation to eliminate $K^{ij} K_{ij}$. We thus obtain the following result for the combination considered previously:
\be
T^i_i = - 4 \pi G \left ( T^{ij} T_{ij}  - \frac{1}{D} (T^i_i)^2 \right ) - \frac{1}{32 \pi G} \dot{A}_i \dot{A}^i
\ee
Now let us consider how this can be interpreted in terms of a field theory deformation. 

By construction, as we take the limit to the conformal boundary, the term of appropriate dilatation weight in $T_{ij}$ becomes the expectation value of the dual stress energy tensor. The (renormalized) expectation value of the dual current ${\cal J}^i$ is similarly related to a specific term in the expansion of the canonical momentum of the gauge field into dilatation eigenfunctions. More specifically, 
\be
\langle {\cal J}^{i} \rangle = \frac{1}{16 \pi G} \pi^i_ {(d)} \label{j-rel}
\ee
where $\pi^i = \dot{A}^i$ and  $\pi^i_{(d)}$ is the coefficient of the $e^{- d r}$ term in $\pi^i$. Note that in general the relationship between $\langle {\cal J}^i \rangle $ and terms in the asymptotic expansion of $\dot{A}^i$ is more complicated. Here the assumed symmetry, i.e. the bulk metric and field strength depend only on the radius, ensures the simple relation between $\langle {\cal J}^i \rangle$ and the coefficient of the $e^{-dr}$ term in the asymptotic expansion.

Let us consider first the case in which there is no source term for the current in the dual field theory. 
The symmetry then guarantees that the asymptotic expansion of the gauge field is 
\be
A^i = A^i_{(d)} e^{-dr} + \cdots
\ee
and thus 
\be
\dot{A}^{i} =  - d e^{-dr} {A}^i_{ (d)}  + \cdots = 16 \pi G e^{- d r} \langle {\cal J}^i \rangle + \cdots
\ee
We can similarly expand
\be
A_i = A_{i (D-1)} e^{- (D-1) r} + \cdots = \gamma_{(0) ij} A^{j}_{(d)} e^{- (D-1) r}  + \cdots
\ee
where $\gamma_{(0) ij}$ is the background metric for the dual field theory (i.e. it is independent of $r$). Thus
\be
\dot{A}_i = - (D-1)  \gamma_{(0) ij} A^{i}_{(d)} e^{- (D-1) r}  + \cdots = 16 \pi G \frac{(D-1)}{d} \gamma_{(0) ij } \langle {\cal J}^j \rangle  e^{-(D-1) r}.
\ee
Note that it is not the case that $\dot{A}_i = \gamma_{ij} \dot{A}^j$ as $\dot{A}_i = \partial_r ( \gamma_{ij} A^{j})$ and $\partial_r (\gamma_{ij}) \neq 0$. 
We can rewrite the trace identity in this case as 
\be
T^i_i = - \lambda  \left ( T^{ij} T_{ij}  - \frac{1}{D} (T^i_i)^2  + \frac{2 (D-1)}{d} e^{- 2 D r} \gamma_{(0) ij} \langle {\cal J}^i \rangle \langle 
{\cal J}^j \rangle + \cdots \right ). \label{tjid}
\ee
where we use the relation $\lambda = 4 \pi G$. This relation suggests that the gauge field contribution is related to a deformation of the field theory involving the square of the symmetry current. 

\bigskip

Now let us return to the general case. In the absence of the gauge field, we interpret $\gamma_{ij}$ as the background metric for the field theory and 
$T_{ij}$ as the operator. While $T_{ij}$ is essentially the canonical momentum dual to $\gamma_{ij}$, we include the counterterm proportional to $\gamma_{ij}$ in the definition of $T_{ij}$ so that the appropriate dilatation eigenfunction in $T_{ij}$ gives the dual CFT stress tensor. Mirroring the same construction for the gauge field, we will define 
\be
{\cal J}^i = \frac{\dot{A}^i}{16 \pi G}
\ee
Now
\be
\dot{A}_i = 2 K_{ij} A^j + \gamma_{ij} \dot{A}^i
\ee
which we can rewrite using 
\be
K_{ij} = 8 \pi G \left ( T_{ij} - \frac{1}{D} T^k_k \gamma_{ij} \right ) + \gamma_{ij}
\ee
giving
\be
\dot{A}^i \dot{A}_i = 16 \lambda^2 \left (  A^j {\cal J}^i (T_{ij} - \frac{1}{D} T^{k}_k \gamma_{ij}) + {\cal J}^i {\cal J}_i \right ) + 8 \lambda A_i {\cal J}^i
\ee
and thus 
\be
T^{i}_i + A_i {\cal J}^i = - \lambda  \left ( T^{ij} T_{ij}  - \frac{1}{D} (T^i_i)^2  +  2  {\cal J}^i {\cal J}_i + 2 A^j {\cal J}^i (T_{ij} - \frac{1}{D} T^k_k \gamma_{ij} )\right ). \label{defx}
\ee
Note that the left hand side of this expression is the standard relation for a CFT, deformed by a source $A_i$ for a current ${\cal J}^i$. 

\subsection{Field theory interpretation}

We now turn to the field theory interpretation. Consider the Ward identity \eqref{defx} on a flat background, with zero direct source $A_i$ for the current i.e.
\be
T^{i}_{i} = - \lambda  \left ( T^{ij} T_{ij}  - \frac{1}{D} (T^i_i)^2  +  2  {\cal J}^i {\cal J}_i \right ).
\ee
To reproduce this Ward identity
we are led to consider the following deformation of a conformal field theory
\be
\lambda \int d^{d} x \left ( T^{ij} T_{ij} - \frac{1}{D} (T^i_i)^2 \right ) + \frac{\lambda}{R^2}  \int d^d x J^i J_i
\ee
where here $J_i$ is a conserved current in the field theory. The coupling of the second term follows on dimensional grounds: the canonical dimension of the current is $D$ and here we denote the volume of the $D$ spatial directions as $V = R^D$ i.e. $R$ is the spatial scale. We will discuss below the relationship between $J_i$ and the current ${\cal J}_i$  defined on the gravity side. 

Following the earlier discussion, let us now consider eigenstates $| n \rangle $ of the energy momentum tensor that are also charge eigenstates i.e.
\be
\langle n | J_\tau | n \rangle = i \frac{q_n}{V}
\ee
where $q_n$ is dimensionless and $V$ is as above the spatial volume. (The factor of $i$ originates from the Euclidean signature.)
Then
\begin{eqnarray}
\langle n | J_\tau J^\tau | n \rangle &=& \sum_m e^{(E_n - E_m) \tau} e^{i (P^a_n - P^a_m) x^a} \langle n | J_{\tau} | m \rangle \langle m | J^{\tau} | n \rangle \\
&=& - \frac{q_n^2}{V^2}, \nonumber
\end{eqnarray}
where we implicitly use the orthonormality of eigenstates of energy and charge. 

The dimensionless energy $\epsilon_n (\alpha, q_n)$ then satisfies the differential equation 
\be
\partial_{\alpha} \epsilon_n = \left ( 1 + \frac{1}{D} \right ) \left ( \epsilon_n^2 + 2 \alpha \epsilon_n \partial_{\alpha} \epsilon_n \right )- q_n^2 
\ee
with the final term originating from the differentation of  the coupling of the current deformation with respect to $\lambda$. We can reinstate a generic normalization for the energy at $\alpha = 0$ as before, giving
\be
\partial_{\alpha} \tilde{\epsilon}_n =\frac{1}{2 C} \left ( \tilde{\epsilon}_n^2 + 2 \alpha \tilde{\epsilon}_n \partial_{\alpha} \tilde{\epsilon}_n \right ) - \frac{2 C d}{D} q_n^2. \label{ft1}
\ee

\subsection{Charged static black branes}

We consider charged static black brane solutions of the type \eqref{static1}. Using the expressions for the curvature \eqref{curv}, it is straightforward to show that the charged black brane solutions for which
\be
f(\rho) = \left ( 1  - \frac{\lambda M}{\rho^{d}} + \frac{Q^2}{\rho^{2D}} \right )
\ee
satisfy Einstein's equations with a (Lorentzian) gauge field such that
\be
(\partial_{\rho} A_t)^2 =  2 D (D-1) \frac{Q^2}{\rho^{2D}} 
\ee
and hence
\be
A_t = \pm \frac{\sqrt{2 D} Q  }{\sqrt{D -1} \rho^{{D-1}}},
\ee
which from \eqref{j-rel} implies that the (renormalized) expectation value of the CFT current is 
\be
\langle {\cal J}_t \rangle = \pm \frac{d \sqrt{2 D}}{4 \lambda \sqrt{D-1}} Q
\ee
according to the usual AdS/CFT dictionary.

The general energy relation \eqref{charg-rel} can be expressed as 
\be
\epsilon(\alpha) = \frac{C}{ \alpha} \left ( 1 -  f(\alpha) \right )
\ee
and
\be
\partial_{\alpha } {\epsilon} - \frac{1}{2 C} {\epsilon} \left ( {\epsilon} + 2 \alpha \partial_{\alpha} \epsilon  \right ) = - \frac{C}{2 \alpha^2} \left (1 - f^2 + \alpha \partial_\alpha (f^2) \right ). \label{ode3}
\ee
For the charged black brane, we can write 
\be
f(\alpha) =  \left ( 1 - \alpha M + {\cal Q}^2 \alpha^2 \right )^{\frac{1}{2}}. \label{bb}
\ee
Here as before $M$ is dimensionless, the gravity normalization is 
\be
C = \frac{D V_D}{2}
\ee
and we define
\be
{\cal Q} = \frac{Q \rho}{\lambda},  \label{curlyq}
\ee
for reasons that we will explain below. 
Note that the energy expanded perturbatively in $\alpha$ is
\be
\epsilon = \frac{C}{2} M +  \alpha \frac{C}{8} (M^2 - 4 {\cal Q}^2) + \cdots 
\ee
where the second term is positive for real $f(\alpha)$. 

\bigskip

Inserting \eqref{bb}  into \eqref{ode3} we can see that the energy relation satisfies
\be
\partial_{\alpha } {\epsilon} - \frac{1}{2 C} {\epsilon} \left ( {\epsilon} + 2 \alpha \partial_{\alpha} \epsilon  \right ) = 
- \frac{C}{2} {\cal Q}^2, 
\ee 
which is of the same form as the field theory result \eqref{ft1}. 

\section{Scalar deformations} \label{sec:six}
 
In this section we briefly consider scalar deformations. We begin by coupling gravity to a single scalar field i.e. the bulk action is 
\be
I_M = - \frac{1}{16 \pi G} \int d^{d+1} x \sqrt{g} \left (R - \frac{1}{2} (\partial \Phi)^2 + V(\Phi) \right )
\ee
where $V(\Phi)$ is the potential for the scalar field. For asymptotically AdS spacetimes, the potential has a fixed point such that $V = d(d-1)$, as above. We can always introduce a coordinate system (locally) in which the metric takes the same form as before i.e.
\be
ds^2 = dr^2 + \gamma_{ij}(x,r) dx^i dx^j.
\ee
Using the Gauss-Codazzi relations we can then write the Einstein equations as 
\begin{eqnarray}
K^2 - K^{ij} K_{ij} &=& {\cal R} + V(\Phi) + \frac{1}{2} \dot{\Phi}^2 - \frac{1}{2} \gamma^{ij} \partial_i \Phi \partial_j \Phi \\
\nabla_{i} K^{i}_j - \nabla_i K &=& \frac{1}{2} \dot{\Phi} \partial_j \Phi \nonumber \\
\dot{K}^{i}_j  + K K^{i}_j &=& {\cal R}^{i}_j - \frac{1}{2} \partial^i \Phi \partial_j \Phi  + \frac{1}{(d-1)} V(\Phi) \delta^i_j \nonumber
\end{eqnarray}
where the dots indicate radial derivatives. The scalar field equation is
\be
\ddot{\Phi}  + K \dot{\Phi}  + \Box_i \Phi + V'(\Phi) = 0, 
\ee
where the prime denotes the derivative with respect to $\Phi$ and $\Box_i$ denotes the Laplacian along the radial slices. 

Let us now restrict to configurations in which the scalar field depends only on the radial coordinate, and the curvature of each constant radius slice is zero. In such a situation, the only relevant counterterm for the action is 
\be
I_{\rm ct} =  \frac{1}{8 \pi G} \int d^{d} x \sqrt{h} W (\Phi)
\ee
where the superpotential $W (\Phi)$ is defined via
\be
V (\Phi)= \frac{d}{ (d-1)} W(\Phi)^2  - 2 W'(\Phi)^2.
\ee
In the case of $V = d (d-1)$, we obtain $W =  (d-1)$ and reproduce the result of the first section.

The quasi-local stress energy tensor follows from the action and is given by
\be
T_{ij} = \frac{1}{8 \pi G}  \left ( K_{ij} - K \gamma_{ij} + W \gamma_{ij} \right ),
\ee
where we have used the fact that radial slices have no intrinsic curvature and the scalar field depends only on the radius. The trace is then
\be
T^i_i = \frac{1}{8 \pi G} \left ( d W - (d-1) K \right ).
\ee
Then we can show that the quadratic combination gives
\begin{eqnarray}
 \left ( T^{ij} T_{ij} - \frac{1}{D} (T^i_i)^2 \right ) &=& \frac{1}{64 \pi^2 G^2} \left (  2  W (K - \frac{d}{D} W ) + (2 (W')^2 - \frac{1}{2} \dot{\Phi}^2 ) \right ) \\
 &=& - \frac{W}{4 \pi G D} T^i_i +  \frac{1}{64 \pi^2 G^2}  \left (2 (W')^2 - \frac{1}{2} \dot{\Phi}^2  \right ) \nonumber
 \end{eqnarray}
 Up this point we have not assumed that the metric has Poincar\'{e} symmetry along the radial slices, i.e. it is a flat domain wall. If we do assume this, then the equations of motion can be expressed in first order form implying that 
 \be
\dot{ \Phi} = 2 W'
 \ee
and 
\be
\dot{A} =  \frac{W}{(d-1)}
\ee 
where the metric is expressed as 
\be
ds^2 = dr^2 + e^{2 A(r)} dx^i dx_i.
\ee
For such flat domain walls, the defining relation is then 
\be
T^{i}_{i} = - \frac{4 \pi G D}{W(\Phi)} \left ( T^{ij} T_{ij} - \frac{1}{D} (T^i_i)^2 \right ). 
\ee 
Here implicitly we interpret the Dirichlet data on each slice as $\gamma_{ij}$ and $\Phi$, and the operators are associated with radial derivatives of these quantities. This suggests that the deformation to be considered in this case is the TT deformation, appropriately dressed by scalar couplings. 

\section{Relation to fluid dual to vacuum Einstein gravity} \label{sec:sixa}

In this section we will consider the relation between gravity at finite radius and a putative dual theory, for spacetimes that are not 
asymptotically AdS. The goal is to connect the discussions of TT deformations in this work with previous analyses of the dual field theory. 

In a number of works including \cite{Bredberg:2010ky,Bredberg:2011jq,Compere:2011dx,Compere:2012mt}, a dual fluid description of vacuum Einstein gravity was explored, building on earlier work on fluid/gravity relations \cite{Fouxon:2008tb,Bhattacharyya:2008kq,Eling:2009pb}.
In this context, one fixes a flat metric $\gamma_{ij}$ on a timelike hypersurface $\Sigma_c$ (outside the Rindler/event horizon) and identifies the Brown-York tensor 
\be
T_{ij} = \frac{1}{8 \pi G} \left ( K \gamma_{ij} - K_{ij} \right ) \label{by}
\ee
as the putative stress tensor of the dual theory. Conservation of the Brown-York tensor translates into integrability conditions for the Einstein equations. 

The Hamiltonian constraint on $\Sigma_c$ can be expressed as
\be
K_{ij} K^{ij} - K^2 = 0 \label{ham1}
\ee
and this constraint can immediately be rewritten in terms of the Brown-York tensor as  \cite{Compere:2011dx,Compere:2012mt}
\be
T^{ij} T_{ij} - \frac{1}{(d-1)} T^2 = 0.
\ee
In \cite{Compere:2011dx,Compere:2012mt}, this relation was interpreted as the equation of state for the dual theory. 

Clearly the definition of the stress tensor via \eqref{by} is not unique: one could equally well define the stress tensor as 
\be
T_{ij}^C = \frac{1}{8 \pi G} \left ( K \gamma_{ij} - K_{ij} + C \gamma_{ij}\right ) \label{byc}
\ee
where $C$ is any constant. In the asymptotically AdS setup, we fix the constant $C$ by requiring that the stress tensor approaches the 
renormalised stress tensor at the conformal boundary. In flat spacetime, there is no a priori reason to fix a particular value of $C$ as we do not have a holographic correspondence at infinite radius. For an equilibrium solution, we can write the stress tensor as 
\be
{\cal T}_{ij} = (p + \rho) u_i u_j + p \gamma_{ij}
\ee
where $p$ is the pressure, $\rho$ is the energy density and $u_i$ is the fluid velocity. The effect of the constant in \eqref{byc} is thus
\be
p \rightarrow p + C \qquad \rho \rightarrow \rho - C.
\ee
In the case of flat space in Rindler coordinates, the stress tensor on a hypersurface $\Sigma_c$ indeed takes the perfect fluid form, with  \cite{Compere:2011dx,Compere:2012mt}
\be
p = 4 \pi T_{U} + C \qquad \rho = - C
\ee
where $T_{U}$ is the Unruh temperature. This motivates choosing $-4 \pi T_U  \le C \le 0$, so that the energy density and the pressure are both non-negative. 

The Hamiltonian constraint \eqref{ham1} is clearly independent of the choice of the stress tensor \eqref{byc}. However, the translation of the Hamiltonian constraint into a constraint on the stress tensor does depend on $C$: for $C \neq 0$, we find
\be
T^C = - \frac{4\pi  (d-1) G}{C} \left ( T^{Cij} T^C_{ij} - \frac{1}{(d-1)} (T^C)^2 \right ) + \frac{d C}{16\pi G}
\ee
Thus for non-zero $C$ the relation (as one would expect) takes a form analogous to that for anti-de Sitter spacetimes. 

\section{Conclusions and outlook} \label{sec:seven}

In this paper we have used the Gauss-Codazzi relations within AdS gravity to write a trace relation for the stress tensor at finite radius. 
This relation suggests that the corresponding deformation of the dual $d$-dimensional CFT is as given by \eqref{res}. As discussed in previous works 
\cite{Guica:2017lia,Kraus:2018xrn,Bzowski:2018pcy}, the deformation is modified by the presence of additional fields in the bulk; we can again use the Gauss-Codazzi relation to deduce systematically the form of the deforming operator.

We used static black brane solutions to derive a relation for the energy at finite radius in terms of the mass parameter of the black brane and the effective coupling. The same relation was reproduced from the perspective of states of a CFT deformed by the operator \eqref{res}. It would be straightforward to extend these results to boosted, spinning branes.

The energy relation is not sensitive to the precise definition of the operator \eqref{res} i.e. how one takes the limit  \eqref{limit-com} to define the composite operator. To explore this deformation further, it would clearly be interesting to calculate quantities that are sensitive to this definition, such as correlation functions and entanglement entropy. Correlation functions in the deformed two-dimensional dual theory were explored in \cite{Kraus:2018xrn}. 
Note however that, since in three bulk dimensions gravity has no propagating degrees of freedom, correlation functions of the stress energy tensor already follow directly from the Ward identities in the presence of sources. In the 2d CFT this is well known: one uses conservation of the stress tensor (the diffeomorphism Ward identity) together with the trace Ward identity
\be
T^i_i = \frac{c}{6} {\cal R},
\ee
where $c$ is the central charge. Relations between stress energy tensor two point functions (in flat space) are obtained by differentiating these identities with respect to the metric and then setting the background metric to be flat. The standard CFT two point functions are then obtained by integrating the relations arising from the diffeomorphism Ward identity, substituting the relations obtained from the trace identity. In the deformed 2d theory, the corresponding trace relation is instead
\be
T^i_i = - \lambda \left ( T^{ij} T_{ij} - (T^i_i)^2 \right ) + \frac{c}{6} {\cal R}. \label{deform-trace}
\ee 
Throughout this paper the last term was set to zero: in the bulk we assumed that hypersurfaces of constant radius are flat and in the field theory we correspondingly took the background metric to be flat. Using the Gauss-Codazzi relation \eqref{gc} we can infer that this would is the generalized relation for non-flat hypersurfaces. Since the stress tensor is still conserved, differentiation of this relation with respect to the metric again gives relations for correlation functions, which are equivalent to those derived in  \cite{Kraus:2018xrn}.

For $d > 2$ one can obtain an analogue of \eqref{deform-trace} using the Gauss-Codazzi relation \eqref{gc}, together with the curvature counterterm contributions to the holographic stress tensor \cite{DeHaro2001}. However, as gravity in $d > 3$ has propagating degrees of freedom, one needs to solve the perturbation equations around AdS to obtain correlation functions; one cannot deduce them from manipulations of the Ward identities. From the field theory perspective, one would need to define the (regularisation of the) composite operator ${\cal T}$ to obtain the correlation functions at finite $\lambda$ using conformal perturbation theory. The regularisation of the composite operator would also be needed to understand entanglement entropy in the deformed theory; see discussions of entanglement entropy for a 2d theory closely related to the $T \bar{T}$ deformed theory in \cite{Chakraborty:2018kpr}.

\section*{Acknowledgements}

This work is funded by the STFC grant ST/P000711/1. This project has received funding and support from the European Union's Horizon 2020 research and innovation programme under the Marie Sklodowska-Curie grant agreement No 690575. MMT would like to thank the Kavli Institute for the Physics and Mathematics of the Universe and the Banff International Research Station for hospitality during the completion of this work.

\bibliographystyle{JHEP}

\begin{thebibliography}{10}

\bibitem{Zamolodchikov:2004ce}
A.~B. Zamolodchikov, \emph{{Expectation value of composite field T anti-T in
  two-dimensional quantum field theory}},
  \href{https://arxiv.org/abs/hep-th/0401146}{{\tt hep-th/0401146}}.

\bibitem{McGough:2016lol}
L.~McGough, M.~Mezei and H.~Verlinde, \emph{{Moving the CFT into the bulk with
  $T\bar T$}},  \href{https://arxiv.org/abs/1611.03470}{{\tt 1611.03470}}.

\bibitem{Jensen:2017tnb}
K.~Jensen, \emph{{Locality and anomalies in warped conformal field theory}},
  \href{http://dx.doi.org/10.1007/JHEP12(2017)111}{\emph{JHEP} {\bf 12} (2017)
  111}, [\href{https://arxiv.org/abs/1710.11626}{{\tt 1710.11626}}].

\bibitem{Guica:2017lia}
M.~Guica, \emph{{An integrable Lorentz-breaking deformation of two-dimensional
  CFTs}},  \href{https://arxiv.org/abs/1710.08415}{{\tt 1710.08415}}.

\bibitem{Giribet:2017imm}
G.~Giribet, \emph{{$T\bar{T}$-deformations, AdS/CFT and correlation
  functions}}, \href{http://dx.doi.org/10.1007/JHEP02(2018)114}{\emph{JHEP}
  {\bf 02} (2018) 114}, [\href{https://arxiv.org/abs/1711.02716}{{\tt
  1711.02716}}].

\bibitem{Kraus:2018xrn}
P.~Kraus, J.~Liu and D.~Marolf, \emph{{Cutoff AdS$_3$ versus the $T\bar{T}$
  deformation}},  \href{https://arxiv.org/abs/1801.02714}{{\tt 1801.02714}}.

\bibitem{Cottrell:2018skz}
W.~Cottrell and A.~Hashimoto, \emph{{Comments on $T \bar T$ double trace
  deformations and boundary conditions}},
  \href{https://arxiv.org/abs/1801.09708}{{\tt 1801.09708}}.

\bibitem{Aharony:2018vux}
O.~Aharony and T.~Vaknin, \emph{{The TT* deformation at large central charge}},
   \href{https://arxiv.org/abs/1803.00100}{{\tt 1803.00100}}.

\bibitem{Bzowski:2018pcy}
A.~Bzowski and M.~Guica, \emph{{The holographic interpretation of $J \bar
  T$-deformed CFTs}},  \href{https://arxiv.org/abs/1803.09753}{{\tt
  1803.09753}}.

\bibitem{Dubovsky:2018bmo}
S.~Dubovsky, V.~Gorbenko and G.~Hernandez-Chifflet, \emph{{$T\bar{T}$ Partition
  Function from Topological Gravity}},
  \href{https://arxiv.org/abs/1805.07386}{{\tt 1805.07386}}.

\bibitem{Chakraborty:2018kpr}
S.~Chakraborty, A.~Giveon, N.~Itzhaki and D.~Kutasov, \emph{{Entanglement
  Beyond $\rm AdS$}},  \href{https://arxiv.org/abs/1805.06286}{{\tt
  1805.06286}}.

\bibitem{deBoer:1999tgo}
J.~de~Boer, E.~P. Verlinde and H.~L. Verlinde, \emph{{On the holographic
  renormalization group}},
  \href{http://dx.doi.org/10.1088/1126-6708/2000/08/003}{\emph{JHEP} {\bf 08}
  (2000) 003}, [\href{https://arxiv.org/abs/hep-th/9912012}{{\tt
  hep-th/9912012}}].

\bibitem{Papadimitriou2004}
I.~Papadimitriou and K.~Skenderis, \emph{{AdS/CFT correspondence and
  Geometry}},  \href{https://arxiv.org/abs/0404176}{{\tt 0404176}}.

\bibitem{Heemskerk:2009pn}
I.~Heemskerk, J.~Penedones, J.~Polchinski and J.~Sully, \emph{{Holography from
  Conformal Field Theory}},
  \href{http://dx.doi.org/10.1088/1126-6708/2009/10/079}{\emph{JHEP} {\bf 10}
  (2009) 079}, [\href{https://arxiv.org/abs/0907.0151}{{\tt 0907.0151}}].

\bibitem{Heemskerk:2010hk}
I.~Heemskerk and J.~Polchinski, \emph{{Holographic and Wilsonian
  Renormalization Groups}},
  \href{http://dx.doi.org/10.1007/JHEP06(2011)031}{\emph{JHEP} {\bf 06} (2011)
  031}, [\href{https://arxiv.org/abs/1010.1264}{{\tt 1010.1264}}].

\bibitem{Brattan:2011my}
D.~Brattan, J.~Camps, R.~Loganayagam and M.~Rangamani, \emph{{CFT dual of the
  AdS Dirichlet problem : Fluid/Gravity on cut-off surfaces}},
  \href{http://dx.doi.org/10.1007/JHEP12(2011)090}{\emph{JHEP} {\bf 12} (2011)
  090}, [\href{https://arxiv.org/abs/1106.2577}{{\tt 1106.2577}}].

\bibitem{Bredberg:2010ky}
I.~Bredberg, C.~Keeler, V.~Lysov and A.~Strominger, \emph{{Wilsonian Approach
  to Fluid/Gravity Duality}},
  \href{http://dx.doi.org/10.1007/JHEP03(2011)141}{\emph{JHEP} {\bf 03} (2011)
  141}, [\href{https://arxiv.org/abs/1006.1902}{{\tt 1006.1902}}].

\bibitem{Bredberg:2011jq}
I.~Bredberg, C.~Keeler, V.~Lysov and A.~Strominger, \emph{{From Navier-Stokes
  To Einstein}}, \href{http://dx.doi.org/10.1007/JHEP07(2012)146}{\emph{JHEP}
  {\bf 07} (2012) 146}, [\href{https://arxiv.org/abs/1101.2451}{{\tt
  1101.2451}}].

\bibitem{Compere:2011dx}
G.~Compere, P.~McFadden, K.~Skenderis and M.~Taylor, \emph{{The Holographic
  fluid dual to vacuum Einstein gravity}},
  \href{http://dx.doi.org/10.1007/JHEP07(2011)050}{\emph{JHEP} {\bf 07} (2011)
  050}, [\href{https://arxiv.org/abs/1103.3022}{{\tt 1103.3022}}].

\bibitem{Compere:2012mt}
G.~Compere, P.~McFadden, K.~Skenderis and M.~Taylor, \emph{{The relativistic
  fluid dual to vacuum Einstein gravity}},
  \href{http://dx.doi.org/10.1007/JHEP03(2012)076}{\emph{JHEP} {\bf 03} (2012)
  076}, [\href{https://arxiv.org/abs/1201.2678}{{\tt 1201.2678}}].

\bibitem{Cardy:2018sdv}
J.~Cardy, \emph{{The $T\overline T$ deformation of quantum field theory as a
  stochastic process}},  \href{https://arxiv.org/abs/1801.06895}{{\tt
  1801.06895}}.

\bibitem{Bonelli:2018kik}
G.~Bonelli, N.~Doroud and M.~Zhu, \emph{{$T\bar T$-deformations in closed
  form}},  \href{https://arxiv.org/abs/1804.10967}{{\tt 1804.10967}}.

\bibitem{DeHaro2001}
S.~de~Haro, K.~Skenderis and S.~N. Solodukhin, \emph{{Holographic
  Reconstruction of Spacetime and Renormalization in the AdS/CFT
  Correspondence}},
  \href{http://dx.doi.org/10.1007/s002200100381}{\emph{Communications in
  Mathematical Physics} {\bf 217} (mar, 2001) 595--622},
  [\href{https://arxiv.org/abs/0002230}{{\tt 0002230}}].

\bibitem{Balasubramanian1999a}
V.~Balasubramanian and P.~Kraus, \emph{{A Stress Tensor for Anti-de Sitter
  Gravity}},
  \href{http://dx.doi.org/10.1007/s002200050764}{\emph{Communications in
  Mathematical Physics} {\bf 208} (dec, 1999) 413--428},
  [\href{https://arxiv.org/abs/9902121}{{\tt 9902121}}].

\bibitem{Smirnov:2016lqw}
F.~A. Smirnov and A.~B. Zamolodchikov, \emph{{On space of integrable quantum
  field theories}},
  \href{http://dx.doi.org/10.1016/j.nuclphysb.2016.12.014}{\emph{Nucl. Phys.}
  {\bf B915} (2017) 363--383}, [\href{https://arxiv.org/abs/1608.05499}{{\tt
  1608.05499}}].

\bibitem{Fouxon:2008tb}
I.~Fouxon and Y.~Oz, \emph{{Conformal Field Theory as Microscopic Dynamics of
  Incompressible Euler and Navier-Stokes Equations}},
  \href{http://dx.doi.org/10.1103/PhysRevLett.101.261602}{\emph{Phys. Rev.
  Lett.} {\bf 101} (2008) 261602}, [\href{https://arxiv.org/abs/0809.4512}{{\tt
  0809.4512}}].

\bibitem{Bhattacharyya:2008kq}
S.~Bhattacharyya, S.~Minwalla and S.~R. Wadia, \emph{{The Incompressible
  Non-Relativistic Navier-Stokes Equation from Gravity}},
  \href{http://dx.doi.org/10.1088/1126-6708/2009/08/059}{\emph{JHEP} {\bf 08}
  (2009) 059}, [\href{https://arxiv.org/abs/0810.1545}{{\tt 0810.1545}}].

\bibitem{Eling:2009pb}
C.~Eling, I.~Fouxon and Y.~Oz, \emph{{The Incompressible Navier-Stokes
  Equations From Membrane Dynamics}},
  \href{http://dx.doi.org/10.1016/j.physletb.2009.09.028}{\emph{Phys. Lett.}
  {\bf B680} (2009) 496--499}, [\href{https://arxiv.org/abs/0905.3638}{{\tt
  0905.3638}}].

\end{thebibliography}

\providecommand{\href}[2]{#2}\begingroup\raggedright\endgroup

\end{document}